\documentclass[twocolumn,superscriptaddress,amsmath,amssymb,showpacs,prb]{revtex4-2}

\usepackage{graphicx}
\usepackage{hyperref}
\usepackage{color}
\renewcommand{\phi}{\varphi}

\begin{document}

\title{Melting Temperature of Iron Under the Earth’s Inner Core Condition from Deep Machine Learning}

\author{Fulun Wu}
	\affiliation{Department of Physics, Xiamen University, Xiamen 361005, China}
\author{Shunqing Wu}
	\affiliation{Department of Physics, Xiamen University, Xiamen 361005, China}
\author{Cai-Zhuang Wang}
    \affiliation{Department of Physics, Iowa State University, Ames, IA 50011, USA}
\author{Kai-Ming Ho}
    \affiliation{Department of Physics, Iowa State University, Ames, IA 50011, USA}
\author{Renata M. Wentzcovitch}
	\affiliation{Department of Applied Physics and Applied Mathematics, Columbia University, New York, NY 10027, USA}
	\affiliation{Department of Earth and Environmental Sciences, Columbia University, New York, NY 10027, USA}
	\affiliation{Lamont–Doherty Earth Observatory, Columbia University, Palisades, NY 10964, USA}
\author{Yang Sun}
	\affiliation{Department of Physics, Xiamen University, Xiamen 361005, China}

\begin{abstract}

Constraining the melting temperature of iron under Earth's inner core conditions is crucial for understanding core dynamics and planetary evolution. Here, we develop a deep potential (DP) model for iron that explicitly incorporates electronic entropy contributions governing thermodynamics under Earth’s core conditions. Extensive benchmarking demonstrates the DP's high fidelity across relevant iron phases and extreme pressure and temperature conditions. Through thermodynamic integration and direct solid-liquid coexistence simulations, the DP predicts melting temperatures for iron at the inner core boundary, consistent with previous \textit{ab initio} results. This resolves the previous discrepancy of iron’s melting temperature at ICB between the DP model and \textit{ab initio} calculation and suggests the crucial contribution of electronic entropy. Our work provides insights into machine learning melting behavior of iron under core conditions and provides the basis for future development of binary or ternary DP models for iron and other elements in the core.

\end{abstract}

\maketitle

\section{Introduction}
The Earth's core plays a critical role in the thermal and compositional evolution of our planet \cite{1,2}. The core contains a solid inner core and a liquid outer, which form a solid-liquid coexistence (SLC) at the inner-core boundary (ICB). Both inner and outer cores are primarily made of iron. It’s generally believed iron has a hexagonal close-packed (hcp) phase under inner core conditions  \cite{3,4}, while the body-centered cubic (bcc) phase is also suggested to be relevant for the inner core structure \cite{5,6,7}. The melting temperature of iron, at which the solid-liquid equilibria form, is a key factor for estimating the temperature of the Earth’s center and the solidification process of the inner core \cite{8,9,10,11,12,13,14,15,16,17,18}. Despite its importance, the melting temperature of iron in the inner core was not well constrained. Experimental measurements reported different results ranging from 4,850 K to 7,600 K due to the difficulties in generating extreme conditions and detecting melts  \cite{19}. The recent experiments still have uncertainties of ~500 K for the melting temperature measurement at ICB conditions  \cite{16,20}. 

Another avenue for exploring the extreme pressure-temperature conditions within Earth's interiors is through computer simulations. These simulations offer a valuable alternative to experimental methods, allowing us to probe the properties of iron under core conditions. Depending on the modeling approach used to describe interatomic interactions, these simulations can be categorized as classical or \textit{ab initio} simulations. Classical molecular dynamics (CMD) simulations employ semi-empirical potentials to represent interatomic interactions. Due to the efficiency of these potentials, CMD can simulate atomic structures at large length scales and over significant time scales, such as millions of atoms over nanoseconds. This allows the direct simulation of solid-liquid equilibria at the ICB conditions so that the melting temperature can be directly extracted  \cite{21}. However, the limitation of CMD is that the simulation results highly depend on the accuracy of the employed semi-empirical potential. Different semi-empirical potential simulations have yielded varying melting temperatures for iron at inner core pressures. For instance, in 2000, Laio et al.  \cite{8} suggested a low iron melting point of $\sim$5,400 K at ICB, while Belonoshko et al. reported a high melting temperature of $\sim$7,100 K for hcp at the ICB \cite{22}. Recently, Davies et al. \cite{23} reported a melting temperature of 6,215 K for hcp Fe at 323 GPa, a pressure close to the ICB, while Sun et al. obtained 5,860 K at the same pressure with another semi-empirical potential \cite{24}. Using a different potential, Belonoshko showed the hcp melting temperature of $\sim$6,400 K at ICB \cite{25,26}. 

Compared to CMD, the \textit{ab initio} molecular dynamics (AIMD) simulations provide more accurate descriptions of the interatomic interaction based on the first-principles electronic structure calculations with density functional theory (DFT)  \cite{27}. However, due to the time and length scale limitations of DFT, direct SLC simulations with AIMD are usually computationally expensive and involve large uncertainty \cite{28}. The lowest uncertainty of iron’s melting temperature from \textit{ab initio} SLC was achieved by Alfè using 980 atoms for hcp phases, resulting in 6,200$\pm$150K  \cite{29}. The free energy approach is more widely used to measure the melting temperature from the AIMD simulations. It is based on the explicit calculation of the Gibbs free energy difference between solid and liquid phases, usually involving thermodynamic integration (TI). TI provides the free energy difference between the target and reference systems for which the absolute free energy is known a priori. Depending on the different reference systems, the technical details of TI can be different. For instance, Alfè et al. employed an inverse-power system as the liquid reference and harmonic crystal as the solid reference and obtained the melting temperature of hcp iron as 6,350$\pm$300K at 330 GPa (6290 K at 323 GPa) \cite{30}. Alfè et al. also used a semi-empirical embedded-atom model as the reference states for solid and liquid iron and obtained the melting temperature as 6,250$\pm$100K at 323 GPa \cite{31}. Sun et al. used the Weeks-Chandler-Andersen gas model for the liquid reference in the TI and a phonon quasiparticle method to compute the free energy of the hcp phase, which resulted in a melting temperature of 6,170$\pm$200K at 330 GPa \cite{32}. Recently, Sun et al. also demonstrated that the embedded-atom model is an efficient reference state for TI calculations and reported a melting temperature of 6,357$\pm$45K at 323 GPa \cite{33}. González-Cataldo and Militzer employed a classical pair potential in TI and reported a melting temperature of 6,523$\pm$8K at 330 GPa \cite{34}. Therefore, the melting temperatures of the hcp phase from \textit{ab initio} TI calculations are most consistent within the uncertainties, except that González-Cataldo and Militzer’s results are $\sim$200 K higher than others. It has been frequently suggested that the number of valence electrons considered in the DFT calculation can significantly affect the measurement of melting temperature for iron under inner-core conditions \cite{32,33,34}. It typically requires 16 valence electrons, i.e., $3s^23p^63d^64s^2$, to address the high-pressure effect and converge free energy calculations. The melting temperature can be underestimated with $3d^64s^2$ electrons \cite{32,33} while overestimated with $3p^63d^64s^2$ electrons \cite{34}.

As the pure AIMD simulations remain a heavy burden for computer resources, interatomic potentials developed with machine learning techniques have significantly extended the timescale and length scale of simulation and maintained the \textit{ab initio} accuracy \cite{35,36,37,38}. The algorithms, such as Neural Network Potential (NNP) \cite{39}, Gaussian Approximation Potential (GAP) \cite{40}, on-the-fly Machine Learning Force Field (MLFF) \cite{41}, and Deep Potential (DP) \cite{42} can incorporate large amounts of \textit{ab initio} data to construct direct mappings from atomic structures to forces and energies, thus saving significant amounts of computational time required for \textit{ab initio} calculation. The melting temperatures of iron under inner core conditions have also recently been studied using machine learning potentials. In particular, Zhang et al. developed a GAP model for iron and reported a melting temperature of 6,253$\pm$170K at 330 GPa \cite{43}, consistent with previous \textit{ab initio} results. However, Yuan and Steinle-Neumann developed a DP model and reported iron’s melting temperature as 7,000-7100 K with an uncertainty of 35 K \cite{44} at 330 GPa. The value predicted by the DP model is significantly higher than those obtained from CMD and AIMD calculations, falling within the range of 6000-6400 K, as summarized above. So far, no explanation has been provided for the large discrepancy. In recent studies, DP models have shown high accuracy in simulating complex minerals in the Earth’s interior, as demonstrated by a few groups in the study of bridgmanite \cite{45,46,47,48}, davemaoite \cite{49}, FeSiO melts \cite{50,51}, and $\delta$-AlOOH \cite{52}, etc. Its poor performance in estimating iron’s melting temperature under inner core conditions is alarming, given that a few studies on elemental partitioning rely on the DP method \cite{44}.

This work aims to develop an accurate DP model and determine melting temperatures for iron phases under inner core pressures with free energy calculations and the SLC method. We will include factors relevant to iron's melting temperature under core conditions. By comparing the present DP model and Yuan's DP model, we try to identify key factors leading to the discrepancy with AIMD results and provide new insight into iron melting behavior under Earth’s core conditions.
 
The paper is organized as follows: Section 2 discusses computational methods used for the DP model development and simulation details. Section 3 provides benchmarks of the present DP model and calculations of melting temperature, and discusses the origin of the discrepancy in Yuan's DP model. At the end, Section 4 concludes the paper.

\section{Methods}
\subsection{Deep-learning potential with electronic entropy contribution}
We developed the DP model for iron under inner core conditions based on the smooth edition descriptor $se\_e2\_a$ proposed by Zhang et al. \cite{53,54}. This descriptor integrates both angular and radial information of atomic configurations to encode the local environment of iron within the cutoff radius. The descriptors $\left\{D_1,D_2,\ldots D_i\right\}$ were used to calculate the free energy via a deep neural network. For an electron-ion system the free energy $F$ is defined as the Mermin free energy \cite{55,56},
\begin{equation}
	F=E+{T_{el}S}_{el}
\end{equation}
where $T_{el}$ is the electronic temperature, $E$ is the self-consistent energy from the Kohn-Sham formalism with orbital occupancies $f_{\mathbf{k}n}$ as
\begin{equation}
	f_{\mathbf{k}n}(T_{el})=\left(1+\exp{\frac{\hbar\left(E_{\mathbf{k}n}-E_f\right)}{k_BT_{el}}}\right)^{-1}
\end{equation}
where $E_{\mathbf{k}i}$ is the one-electron energy of an orbital with wavenumber $\mathbf{k}$ and band index $n$, and $E_f$ is the Fermi energy. $S_{el}$ is the electronic entropy, defined by
\begin{equation}
	S_{el}\left(f_{\mathbf{k}n},T_{el}\right)=-k_B\sum_{\mathbf{k},n}\left[\left(1-f_{\mathbf{k}n}\right)\ln{\left(1-f_{\mathbf{k}n}\right)+f_{\mathbf{k}n}\ln{f_{\mathbf{k}n}}}\right].
\end{equation}

The electronic entropy contribution plays a key role in determining the free energy of metals, particularly important for iron under high pressure and temperature conditions \cite{57}. Figure 1 shows the training scheme to incorporate the electronic entropy contribution in the neural network model, which results in an electronic temperature-dependent DP model \cite{58}. As shown in Fig. 1, the local environment surrounding each atom $i$ and electronic temperature $T_{el}$ are input parameters. The total free energy $F$ is obtained by summing up the contributions from all atoms, $F=\sum_{i=1}^{N}F_i$. To preserve physical symmetry, the relative coordinates of atoms were mapped onto generalized coordinates, ${\hat{x}}_{ji}=\frac{s\left(r_{ji}\right)x_{ji}}{r_{ji}}$, using a continuous and differentiable scalar weighting function, $s\left(r_{ji}\right)$. The neighbor atom cutoff radius was set to 6.0 $\AA$. Our embedding and fitting neural networks had three hidden layers with $\{25, 50, 100\}$ and $\{240, 240, 240\}$ neurons, respectively. We randomly initialized the neural network and trained it for 1,000,000 steps using the Adam stochastic gradient descent method \cite{59}. We set the learning rate to decrease exponentially, with the decay step and decay rate being 5,000 and 0.96, respectively. The loss function was defined as a combination of the energy prefactor $p_e$, force prefactor $p_f$, and virial prefactor $p_\xi$ \cite{53} as
\begin{equation}
	L\left(p_e,p_f,p_\xi\right)=\frac{p_e}{N}\Delta A^2+\frac{p_f}{3N}\sum_i|\Delta F_i|^2+\frac{p_\xi}{9N}\parallel\Delta\mathrm{\Xi}\parallel^2
\end{equation}
where the root mean square errors in energy, force, and virial were represented by $\Delta A$, $\Delta F_i$ and $\Delta\mathrm{\Xi}$, respectively. We started with the energy prefactor $p_e$ at 0.2 and gradually increased to 1, while the virial prefactor $p_\xi$ started at 0.01 and gradually increased to 0.1. The force prefactor $p_f$ decreased from 1000 to 1 to achieve the desired accuracy in force prediction.

\begin{figure}
\includegraphics[width=0.49\textwidth]{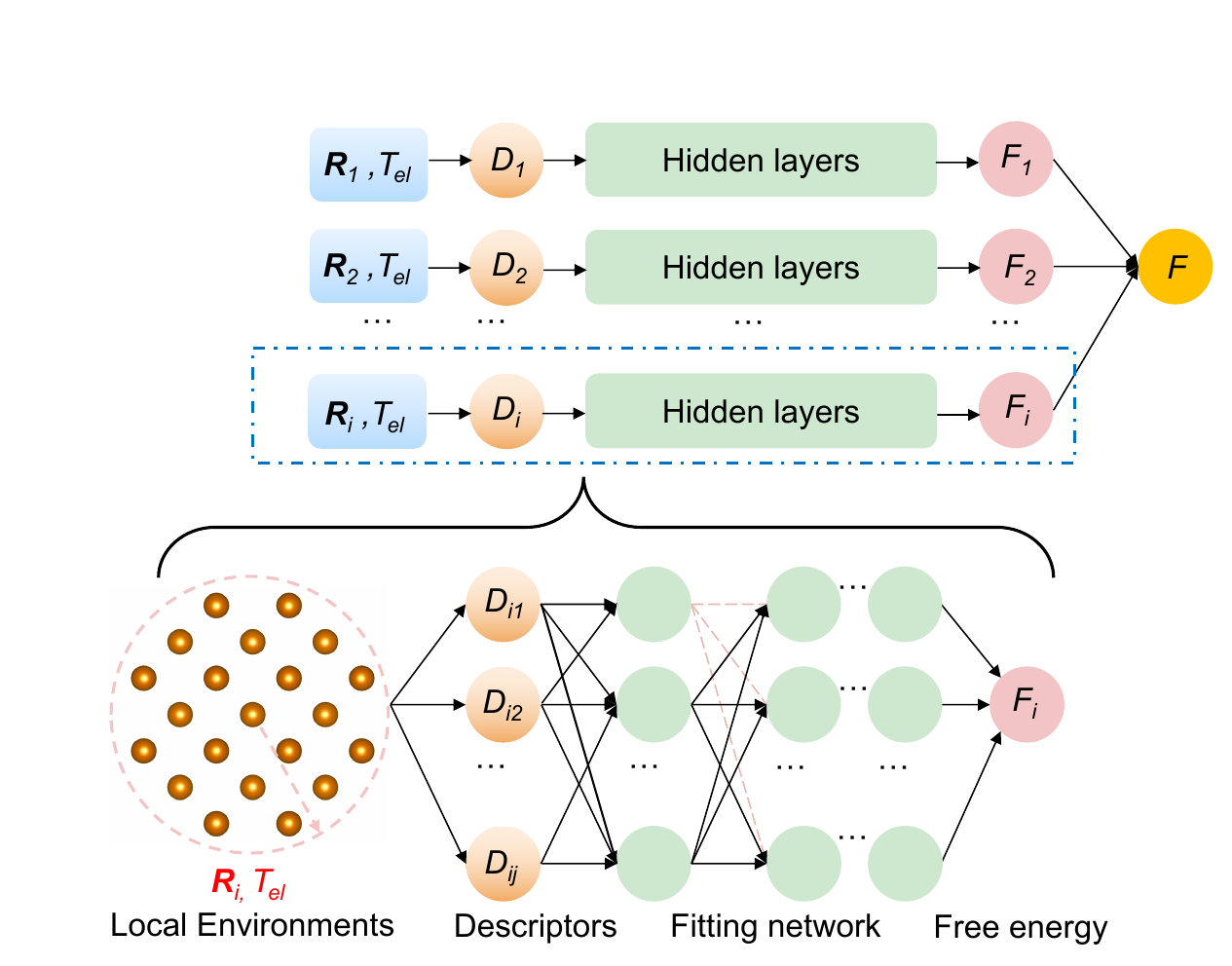}
\caption{The scheme to calculate the atomic free energy of iron from their local environment using a neural network. The atomic local environment $\{R_1,R_2,\ldots R_i\}$ and electronic temperature($T_{el}$) are used as input parameters.}
\end{figure}

\subsection{Density functional theory calculations}
DFT calculations were conducted to prepare training data of iron using the Vienna \textit{Ab-initio} Simulation Package (VASP) \cite{60,61}, which implements the projector-augmented wave (PAW) methodology \cite{62,63}. The exchange-correlation functional was treated with the generalized gradient approximation (GGA) \cite{63} in the form of the Perdew-Burke-Ernzerhof (PBE) formula. PAW potential with 16 valence electrons ($3s^23p^63d^74s^1$) was used for iron. The plane wave energy cutoff was set to 750 eV. The Brillouin zone was sampled using a Monkhorst-Pack scheme with a k-point mesh of $2\times2\times2$. This setting has been shown to achieve the high accuracy necessary for iron under Earth's core conditions \cite{32,33}. The electronic entropy in DFT calculations is described by the Mermin functional \cite{55,56}, with the electronic temperature $T_{el}$ kept the same as the ionic temperature. The DFT calculations were performed with 288 atoms for hcp, 250 atoms for bcc, and 250 for liquid.

\subsection{Thermodynamic integration}
To calculate the free energies with sufficient time and length scales, the TI scheme developed in \cite{33} was employed, which provides a transformation of the Hamiltonian from a classical reference system to the DP system. We used the classical embedded-atom model developed in \cite{24} as the reference state, where the classical free energy difference between liquid and solid was computed, denoted as $\Delta G_C^{L-S}\left(T\right)$. The transformation from classical free energy $\Delta G_C^{L-S}(T)$ to DP free energy $\Delta G_{DP}^{L-S}(T)$ can be obtained by considering the contribution $f_{pV}(T)$ from the equation of state (EoS) difference between DP and classical systems, and the Helmholtz free energy contribution $f_{TI}(T)$ computed by the TI between the liquid and solid phases as
\begin{equation}
	\Delta G_{DP}^{L-S}\left(T\right)=\Delta G_C^{L-S}\left(T\right)+f_{pV}\left(T\right)+f_{TI}\left(T\right)
\end{equation}
\begin{align}
	f_{PV}\left(T\right)=\left[P\left(V_A^L-V_A^S\right)-P\left(V_C^L-V_C^S\right)\right]\\
	-\int_{V_C^L}^{V_{DP}^L}{P_C^L\left(V\right)dV}+\int_{V_C^S}^{V_{DP}^S}{P_C^S\left(V\right)dV}
\end{align}
\begin{align}
		f_{TI}\left(T\right)=\int_{0}^{1}{\langle U_{DP}^L-U_C^L\rangle_{\lambda,NVT}d\lambda}\\
		-\int_{0}^{1}{\langle U_{DP}^S-U_C^S\rangle_{\lambda,NVT}d\lambda}
\end{align}
where $V_{DP}^L$ (or $V_{DP}^S$) and $V_C^L$ (or $V_C^S$) are the equilibrium volumes of the liquid (or solid) at pressure $P$ for DP and classical systems, respectively. $P_C^L(V)$ and $P_C^S(V)$ represent the equation of states of the liquid and solid for the classical system, respectively. $U_{DP}^L$ (or $U_{DP}^S)$ and $U_C^L$ (or $U_C^S$) are the internal energies of the liquid (or solid) for DP and classical systems, respectively. The ensemble average of the internal energy over configurations $\left\langle\cdot\right\rangle_{\lambda,NVT}$ was sampled in the canonical ensemble with the mixed force field $U=\left(1-\lambda\right)U_{DP}+\lambda U_C$. The subscript NVT represents the constant conditions of $(V_{DP}^L, T)$ and $(V_{DP}^S, T)$ in the liquid and solid simulations, respectively. The TI was performed with 2,000 atoms for hcp, bcc, and liquid phases. 

\subsection{Solid-liquid coexisting method}
The solid-liquid coexisting method was also employed to determine the melting point of iron \cite{21}. To prepare the two-phase configurations with a solid-liquid interface, an iron slab with 15,552 atoms was first equilibrated at a target temperature. Then, half of the atoms were fixed while the other half were heated until melting occurred. The liquid part was subsequently cooled to the target temperature and the entire slab was subjected to the target temperature, allowing one phase to grow at the expense of the other over time. The melting temperature was identified as the temperature at which the liquid portion either increased above or decreased below the test value.

\section{Results}
\subsection{Benchmarks of the deep potentials}
The configurations obtained from the \textit{ab initio} data span various pressure and temperature ranges, covering 323-360 GPa and 6300-6700 K. To evaluate the accuracy of the $T_{el}$-dependent DP model, we examined the root mean square errors (RMSEs) of the energies, atomic forces, and pressures for iron and compared them with the corresponding density functional theory (DFT) calculations, as shown in Figure 2(a-c). We found an excellent agreement between the results of the $T_{el}$-dependent DP model and DFT. The RMSEs were approximately 4.6 meV/atom for energies, 0.32 eV/$\AA$ for forces, and 0.47 GPa for pressures. Figure 2(a) also revealed a wide distribution of energies, suggesting a sufficient exploration of complex configuration spaces on the potential energy surface. We further investigated the RMSEs of the DP when the electronic entropy was not specifically included in the training procedure, and compared them with the results obtained from DFT calculations. Figure 2(d-f) clearly illustrates substantial discrepancies in energy and pressure between the DP and DFT outcomes. The discrepancies can be attributed to the inconsistent electronic entropy at different temperatures. This highlights the significance of incorporating the electronic entropy contribution in the training process for iron under inner core conditions to correctly describe the free energy. The force was not affected by the electronic entropies. By inspecting Yuan's training dataset \cite{64}, we found the AIMD simulations in \cite{44} were performed with a fixed electronic temperature $T_{el}$=6000K for different ionic temperatures. Therefore, Yuan's training dataset and DP model failed to describe the electronic entropy contribution correctly.

\begin{figure}
\includegraphics[width=0.49\textwidth]{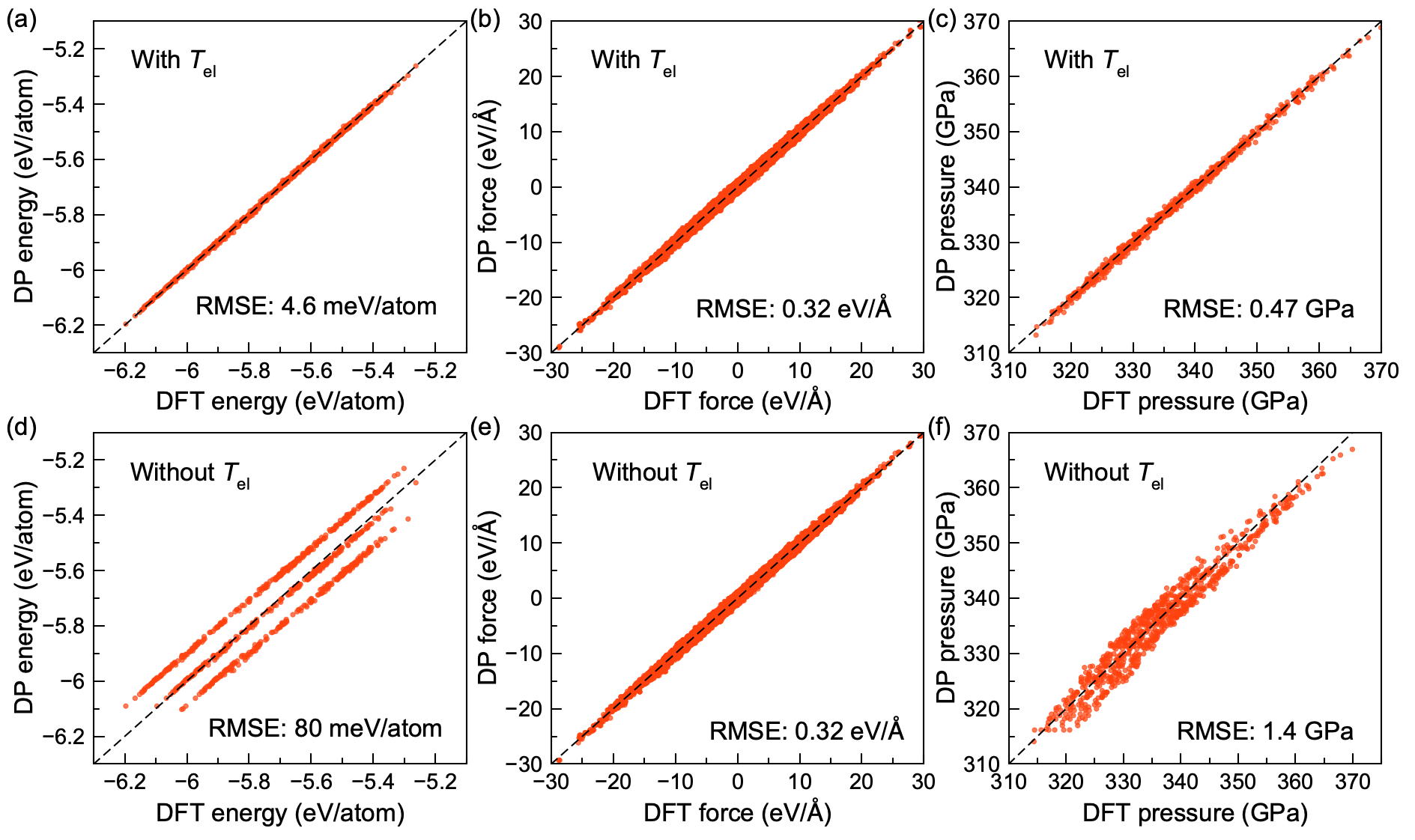}
\caption{Comparisons of (a) energies, (b) atomic forces, and (c) pressures between DFT and the $T_el$-dependent DP model. (d-f) denote the results between DFT and DP trained without the electronic entropy contribution. The black dashed lines are guides for perfect matches.}
\end{figure}

The DP model is further validated by comparing the equation of state (EOS) of iron phases. Figure 3 shows the EOS of bcc, hcp, and liquid iron obtained from DP simulations agree well with recent DFT calculations and experiments \cite{65,66,67}. This further suggests the present DP model can well describe all bcc, hcp and liquid phases under core conditions.

\begin{figure}
\includegraphics[width=0.49\textwidth]{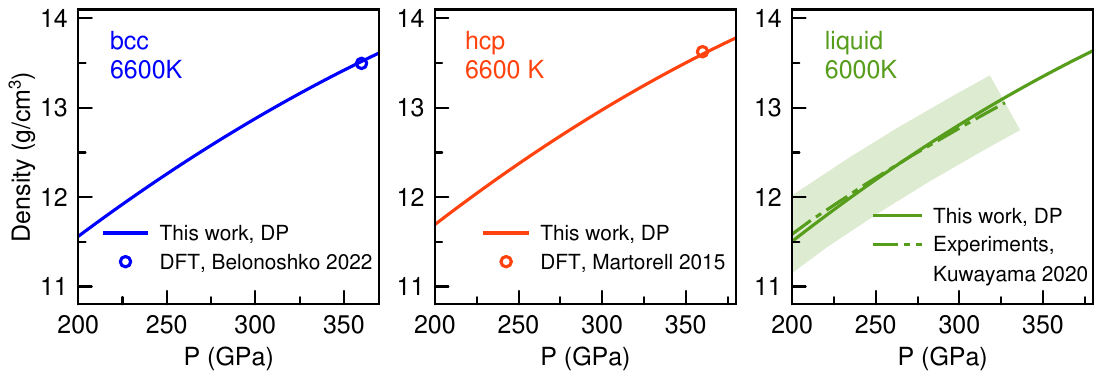}
\caption{Equation of state of bcc, hcp, and liquid iron between DP and DFT \cite{65,66} or experimental results \cite{67}. The shadow in the liquid EoS curve is the experimental uncertainty.}
\end{figure}

\subsection{Melting temperatures}
We first use the free energy calculation to compute the melting temperature for the present DP model. Based on the TI method, we compute the Gibbs free energy difference between solid and liquid, $\Delta G^{L-S}$, under various pressure and temperature conditions. Figure 4 shows the Gibbs free energy difference as a function of temperature at 360 GPa. A negative value of $\Delta G^{L-S}$ implies that the liquid phase is more thermodynamically stable compared to the solid phase, while $\Delta G^{L-S}(T)=0$ corresponds to the melting temperature $T=T_m$. In Fig. 4 (a), the DP-based simulations resulted in melting temperatures of 6656 K for hcp and 6480 K for bcc at 360 GPa. Compared to previous AIMD simulations \cite{33}, $\Delta G^{L-S}$ curves of both hcp and bcc only show a difference of 3-5 meV/atom at the melting points. However, this small energy difference yields a melting temperature $\sim$40 K lower. This suggests the melting temperature is very sensitive to the accuracy of free energy calculations. Nevertheless, the present DP model provides a satisfactory description of the free energy compared to previous AIMD data. In Fig. 4(b), our $T_{el}$-DP model provides a very different melting curve compared to the DP model generated by Yuan and Steinle-Neumann \cite{44} with a difference of more than 600 K.

\begin{figure}
\includegraphics[width=0.49\textwidth]{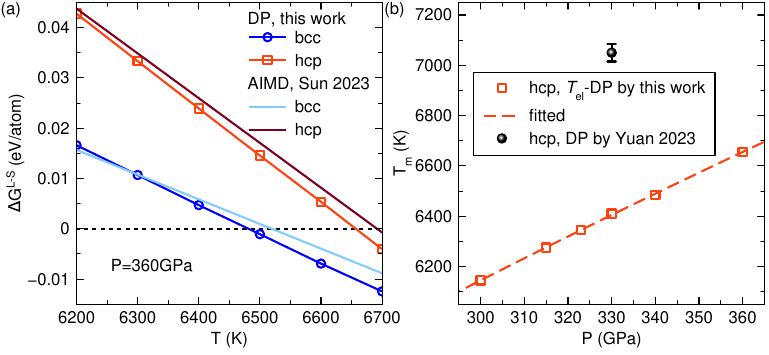}
\caption{(a) The Gibbs free energy difference $\mathrm{\Delta}G_{L-S}$ for bcc and hcp at 360 GPa. The intersection with the dotted line $\mathrm{\Delta}G_{L-S}\ =0$ defines the melting temperature. The DFT results were from \cite{33}. (b) The melting temperatures of the hcp phase. The data from the present DP model is fitted as $T_m=3010+12.07P-0.005410P^2$.}
\end{figure}

Because the melting temperatures in Yuan and Steinle-Neumann \cite{44} were computed by SLC simulations, we also computed the hcp melting temperature by the SLC simulations to examine whether different melting temperature calculation methods could introduce errors. Figure 5 shows the change of hcp populations in a series of SLC simulations at different temperatures at 330 GPa. The increase of the hcp atom number indicates the simulation temperature is below the melting temperature, while the decrease of the hcp atom number indicates the temperature is above the melting temperature. The data in Fig. 5 suggests the melting temperature is $\sim$6420 K at 330 GPa. This is consistent with the free energy results of 6410 K at 330 GPa. Therefore, SLC and free energy calculations do not show a significant difference in melting temperature calculations.

\begin{figure}
\includegraphics[width=0.49\textwidth]{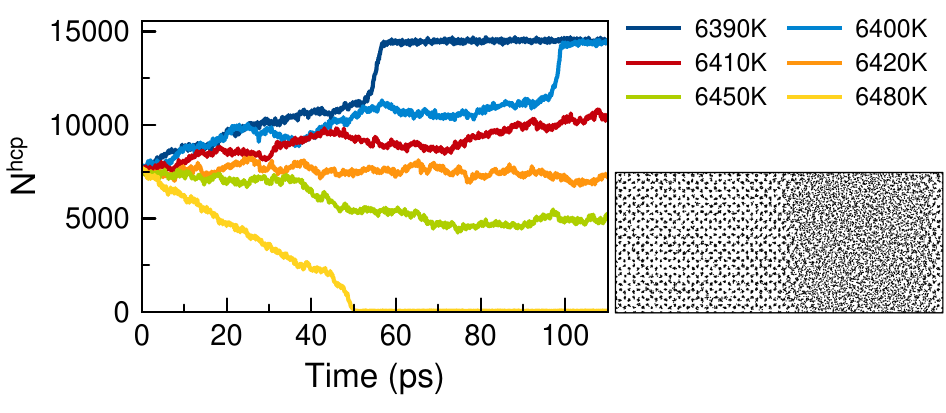}
\caption{The melting temperature at 330 GPa from the SLC method. Different curves show the number of hcp atoms as a function of simulation time at different temperatures. The right panel shows a snapshot from the SLC simulation.}
\end{figure}

\section{Discussions}
We have shown that the DP method can accurately describe the EoS and Gibbs free energies for different iron phases under inner core conditions. Unlike Yuan's results \cite{44}, the melting temperatures from our DP model are very consistent with the ones from AIMD simulations. In many works \cite{37}, DP models were usually trained on the ground-state, Born-Oppenheimer energy surface, which only considers atomic positions but does not take into account the simulation temperature. Therefore, the effect of electronic entropy was disregarded. This treatment may not be significant for systems at low temperatures. However, at the temperature relevant to Earth’s core, the electronic entropy plays a crucial role in the energy and pressures, as shown in Fig. 2. The contribution from electronic entropy was noticed by Belonoshko in the development of classical potential, and was added as an ad hoc correction to Gibbs free energy from CMD simulations \cite{25}. We believe the disregard for electronic entropy contributed to the underestimation of liquid-free energy in Yuan's DP model, resulting in a high hcp phase melting temperature. We also noticed a few differences in the DFT calculations between ours and Yuan's work. We employed the iron’s PAW potential with 16 valence electrons ($3s^23p^63d^64s^2$), while Yuan et al. used one containing 14 valence electrons ($3p^63d^64s^2$) without $3s$ electron contributions \cite{44}. The Brillouin zone was sampled with a $2\times2\times2$ k-point grid in our DFT calculations, whereas a single $\Gamma$ point was used in Yuan's work \cite{44}. The accumulation of these errors resulted in the significant discrepancy in melting temperature computed from \cite{44}. In Fig. 5, we summarize the melting points of the hcp phase at ICB from recent theoretical calculations \cite{16,17,20,23,25,29,30,32,33,34,44,68,69,70,71,72,73} and experiments. We find most data are located in the region of 6,370$\pm$200 K, which can be used as a representative value of iron’s melting points at the ICB.

\begin{figure}
\includegraphics[width=0.49\textwidth]{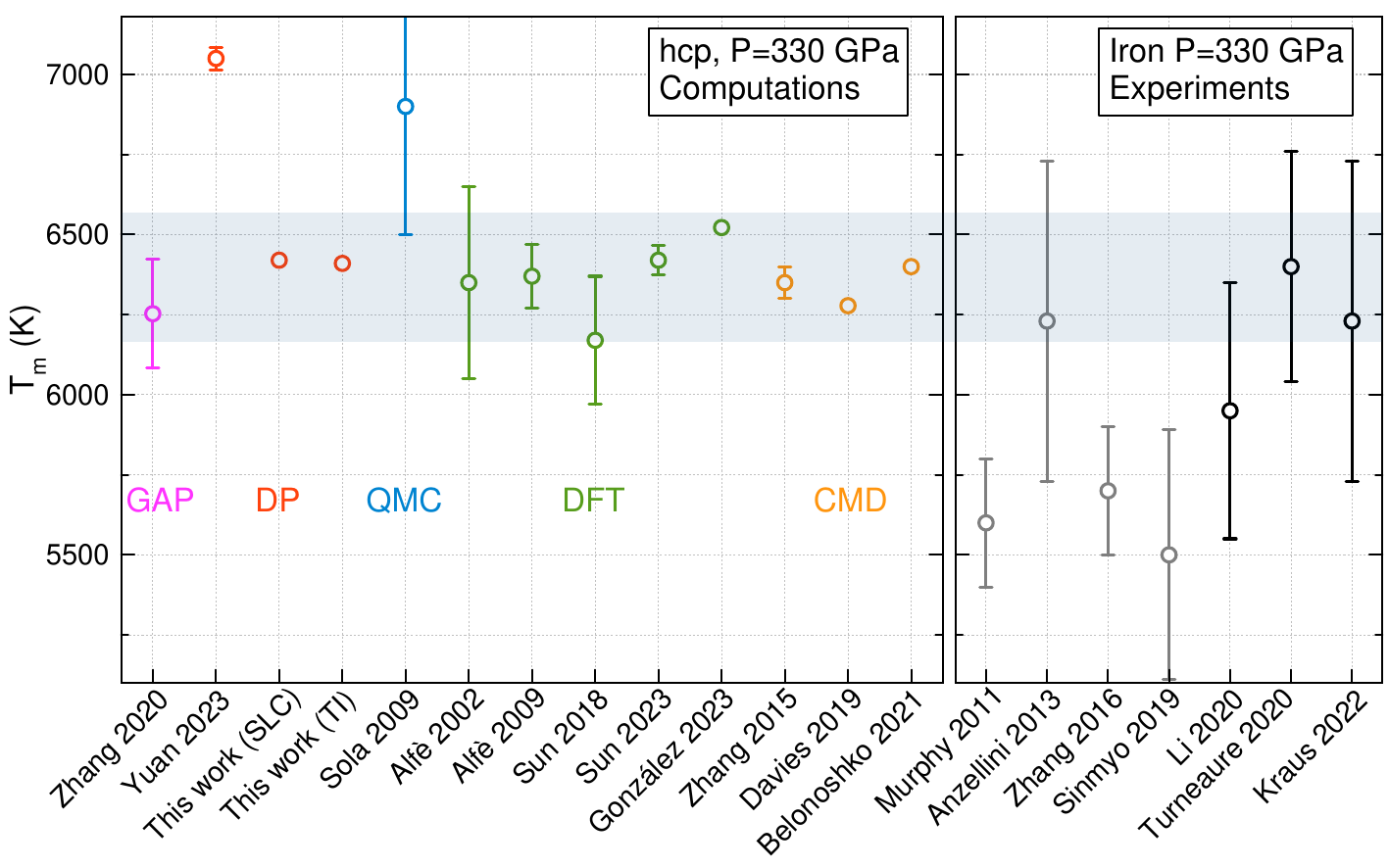}
\caption{Current and previous results on hcp iron’s melting temperatures obtained from calculations and experiments. The shadow region is 6,370$\pm$200K.}
\end{figure}

\section{Conclusion}
In summary, we developed an electronic temperature-dependent DP model for iron under inner core conditions. With thermodynamic integration and solid-liquid coexisting methods, we investigate the melting temperature of iron under Earth's inner core conditions. The DP model accurately reproduced energies, forces, and pressures compared to DFT calculations. It provides the EoS and melting temperatures of hcp iron that are consistent with previous AIMD simulations. We resolved the discrepancy of melting temperature from the previous DP model and showed the importance of including electronic entropy effects in describing iron’s free energy under inner core conditions. The calculations from the present DP model and most previous computational and experimental data suggest the melting temperature of hcp iron in the region of 6,370$\pm$200 K at ICB pressure. Our work provides insights into the machine learning melting behavior of iron under core conditions and provides the basis for future development of binary or ternary DP models for iron and other elements in the core.

\begin{acknowledgments}
Work at Xiamen University was supported by NSFC (Grants No. 42374108 and No. 12374015). R.M.W. acknowledges support from NSF (Grants No. EAR-2000850 and No. EAR-1918126). K.M.H. acknowledges support from NSF Grant No. EAR-1918134. Shaorong Fang and Tianfu Wu from the Information and Network Center of Xiamen University are acknowledged for their help with Graphics Processing Unit (GPU) computing.
\end{acknowledgments}

\bibliographystyle{apsrev4-1}

\end{document}